# Transitioning to human interaction with AI systems:

# New challenges and opportunities for HCI professionals to enable human-centered AI

Wei Xu [a*], Marvin J. Dainoff [b], Liezhong Ge [a],  Zaifeng Gao [c]

[a] Center for Psychological Sciences, Zhejiang University, Hangzhou, China; [b] Dept. of Psychology, Miami University, Oxford, Ohio; [c] Zhejiang University, Department of Psychology and Behavioral Sciences, Hangzhou, China

## Biographies

**Wei Xu** received his Ph.D. in Psychology with specialization in HCI/Human Factors and his M.S. in Computer Science from Miami University in 1997. He is a Professor of HCI/Human Factors at the Center for Psychological Sciences of Zhejiang University, China. His research interests include human-AI interaction, human-computer interaction, and aviation human factors.

**Marvin J. Dainoff** received his Ph.D. in Psychology from University of Rochester in 1969.  He is a Professor Emeritus of Psychology/Human Factors at Miami University. He is a Past President of the Human Factors and Ergonomics Society. His research interests include sociotechnical system solutions for complex systems, human-computer interaction, and workplace ergonomics.

**Liezhong Ge** received his Ph.D. in Psychology with specialization in human factors from Zhejiang University, China, in 1992. He is a Professor of HCI/Human Factors at the Center for Psychological

---
* Address correspondence to Wei Xu, email: weixu6@yahoo.com

Sciences of Zhejiang University. His research interests include human-computer interaction, user experience, and facial recognition.

**Zaifeng Gao** received his Ph.D. in Psychology from Zhejiang University, China, in 2009. He is a Professor of Psychology/Human Factors at the Department of Psychology and Behavioral Sciences, Zhejiang University. His research interests include engineering psychology, autonomous driving, and cognitive psychology.

**Abstract**

While AI has benefited humans, it may also harm humans if not appropriately developed. The priority of current HCI work should focus on transiting from conventional human interaction with non-AI computing systems to interaction with AI systems. We conducted a high-level literature review and a holistic analysis of current work in developing AI systems from an HCI perspective. Our review and analysis highlight the new changes introduced by AI technology and the new challenges that HCI professionals face when applying the human-centered AI (HCAI) approach in the development of AI systems. We also identified seven main issues in human interaction with AI systems, which HCI professionals did not encounter when developing non-AI computing systems. To further enable the implementation of the HCAI approach, we identified new HCI opportunities tied to specific HCAI-driven design goals to guide HCI professionals addressing these new issues. Finally, our assessment of current HCI methods shows the limitations of these methods in support of developing AI systems. We propose the alternative methods that can help overcome these limitations and effectively help HCI professionals apply the HCAI approach to the development of AI systems. We also offer strategic recommendation for HCI professionals to effectively influence the development of AI systems with the HCAI approach, eventually developing HCAI systems.

**Corresponding author:**  Wei Xu; weixu6@yahoo.com

## 1.  Introduction

### 1.1 History is repeating

When the personal computers (PC) first emerged in the 1980s, the development of its applications basically adopted a "technology-centered approach," ignoring the needs of ordinary users. With the popularity of PC, the problems of user experience were gradually exposed. With the initiative primarily driven by human factors, psychology, and computer science, the field of human computer interaction (HCI) came into being (Baecker et al., 1995). As an interdisciplinary field, HCI adopts a "human-centered design" approach to develop computing products meeting user needs. Accordingly, SIG CHI (Special Interest Group on Computer–Human Interaction) of the Association for Computing Machinery (ACM) was established in 1982. The first SIG CHI Annual Conference on Human Factors in Computing Systems was held in 1983 and the main theme of "human-centered design" has been reflected in the annual conferences since 1983 (Grudin, 2005).

History seems to be repeating; we are currently facing a similar challenge as seen in 1980s as we enter the era of artificial intelligence (AI). While AI technology has brought in many benefits to humans, it is having a profound impact on people's work and lives. Unfortunately, the development of AI systems is mainly driven by a "technology-centered design" approach (e.g.,

Shneiderman, 2020a, 2020b; Xu, 2019a; Zheng et al., 2017). Many AI professionals are primarily

dedicated to studying algorithms, rather than providing useful AI systems to meet user needs,

resulting in the failure of many AI systems (Hoffman et al., 2016; Lazer et al., 2014; Lieberman,

2009; Yampolskiy, 2019). Specifically, the AI Incident Database has collected more than 1000 AI

related accidents (McGregor et al., 2021), such as an autonomous car killing a pedestrian, a trading

algorithm causing a market "flash crash" where billions of dollars transfer between parties, and

a facial recognition system causing an innocent person to be arrested. When more and more AI-

based decision-making systems such as enterprises and government services are put into use,

decisions made using biased "thinking" will directly affect people's daily work and life, potentially harming them. AI systems trained with distorted data may produce biased "thinking," easily amplify prejudice and inequality for certain user groups, and even harm individual humans. Like the "dual use" nature of nuclear energy and biochemical technology, the rewards and potential risks brought by AI coexist. Since the development and use of AI is a decentralized global phenomenon, the barriers to entry are relatively low, making it more difficult to control (Li, 2018).

　　As the HCI field that was brought into existence by PC technology 40 years ago, history has brought HCI professionals to a juncture. This time the consequence of ignoring the "human-centered design" philosophy is obviously more severe. Concerning the potential harm to humans from AI technology in the long run, prominent critics have expressed their concern that AI may create a "machine world" and eventually replace humans (Hawking, Musk et al., 2015; Russell et al., 2015). Leading researchers have also raised deep concerns (e.g., Stephanidis, Salvendy et al.,

2019; Shneiderman 2020 a, 2020c; Hancock, 2019; Salmon, Hancock & Carden, 2019; Endsley,

2017, 2018; Lau et al., 2020). Salmon, Hancock & Carden (2019) urged: "the ball is in our court,

but it won't be for much longer… we must take action." Hancock (2019) further described the current boom in the development of AI-based autonomous technologies as "a horse has left the stables" while Salmon (2019) believes that "we find ourselves in a similar situation: chasing a horse that has really started to run wild." Stephanidis, Salvendy et al. (2019) conducted a comprehensive review that identifies seven grand challenges with intelligent interactive systems and proposed the research priorities to address the challenges. In addition, twenty-three scholars from MIT, Stanford, Harvard and other universities jointly published a paper on machine behavior in *Nature* (Rahwan et al., 2019), urging that society should fully understand unique AI-based machine behavior and its impacts to humans and society.

To respond to the challenges, Stanford University established a "Human-Centered AI" research institution: focusing on "ethically aligned design" (Li, 2018; Donahoe, 2018). Shneiderman and Xu have moved forward further proposing a human-centered AI (HCAI) approach (Shneiderman, 2020a, 2020b, 2020c; Xu, 2019a, 2019b; Xu & Ge, 2020). Shneiderman (2020a, 2020d) further proposes a design framework and a governance structure of ethical AI for implementation of HCAI. Others also promoted HCAI different perspective, for instance, Cerejo (2021) proposed an alternative development process to implement HCAI. Overall, the promotion of HCAI is still in its infancy, there is a need to further advance the HCAI approach and the promotion to a wide audience, especially the HCI community as HCAI is rooted in the "human-centered design" philosophy that has been adopted by HCI.

## 1.2 Unique autonomous characteristics of AI technology

We encounter many unique issues and risks introduced by AI technology which do not exist when dealing with non-AI computing systems.  HCI professionals need to fully understand these unique characteristics of AI technology so that we can take effective approaches to address their consequences.

AI technology has brought in unique autonomous characteristics that specifically refers to the ability of an AI-based autonomous system to perform specific tasks independently. AI systems can exhibit unique machine behaviors and evolve to gain certain levels of human-like cognitive/intelligent capabilities, such as, self-executing and self-adaptive abilities; they may successfully operate under certain situations that are possibly not fully anticipated, and the results may not be deterministic (den Broek et al., 2017; Kaber, 2018; Rahwan et al., 2019; Xu & Ge, 2020). These autonomous characteristics are typically used to define the degree of intelligence of an AI system (Watson & Scheidt, 2005).

In contrast, automation represents the typical characteristics of non-AI computing systems. Automation is the ability of a system to perform well-defined tasks and to produce deterministic results, typically relying on a fixed set of rules or algorithms based on mechanical or digital computing technology. Automation cannot perform the tasks that were not designed for; in such cases, the operators must manually take over the automated system (Kaber, 2018). For example, the autopilot on a civil aircraft flight deck can carry out certain flight tasks previously undertaken by pilots, moving them away from their traditional role of directly controlling the aircraft to a more supervisory role managing the airborne automation. However, in abnormal situations which the autopilot was not designed for, pilots must immediately intervene to take over manual control of the aircraft (Xu, 2007).

Table 1 compares the characteristics between non-AI based automated systems and AI-based autonomous systems from an HCI perspective (Kaber, 2018; Xu, 2021; den Broek et al., 2017; Bansal et al., 2019a). As Table 1 shows, the fundamental difference in capabilities between the two is that autonomous systems can be built with certain levels of human-like cognitive/intelligent abilities (Kaber, 2018; Xu, 2021). On the other hand, both automated and autonomous systems require human intervention in operations for safety. These differences and similarities between the two are of great significance for HCI solutions in developing AI systems.

**Table 1 Comparative analysis between non-AI-based automation and AI-based autonomy**
(Based on Kaber, 2018; den Broek et al., 2017; Bansal et al., 2019a; Rahwan et al., 2019; Xu, 2021)

| Characteristics | Non-AI Based Automated Systems (With varied levels of automation) | AI-Based Autonomous Systems (With varied degrees of autonomy) |
|---|---|---|
|  | **Examples**: conventional office software, washing machine, elevator, automated manufacturing lines | **Examples:** smart speakers, intelligent decision systems, autonomous vehicles, intelligent robots |
| **Human-like sensing ability** | Limited | Yes (With advanced technologies) |
| **Human-like cognitive abilities** (pattern recognition, learning, reasoning, etc.) | No | Yes (Abilities vary across design scenarios) |
| **Human-like self-executing ability** | No (Require human manual activation and intervention according to predefined rules | Yes (Perform operations independently in specific situations: abilities vary across scenarios) |
| **Human-like self-adaptive ability to unpredictable environments** | No | Yes (Abilities vary across scenarios) |

| Operation outcomes | Deterministic | Non-deterministic, unexpected |
| --- | --- | --- |
| **Human intervention** | Human intervention is required. Human must be the ultimate decision maker | |

It should be noted that some characteristics of AI systems as listed in Table 1 depend on the high degree of autonomy to be driven by future AI technology, which may not exist today and may be available as predicted (e.g., Bansal et al., 2019b; Demir et al., 2018a). The distinction between automation technology versus AI-based autonomy technology has been debated (e.g., O'Neill et al., 2020; Kaber, 2018). Kaber (2018) believes that recent human-automation interaction research has confused concepts of automated systems and autonomous systems, which has led to inappropriate expectation for design and misdirected criticism of design methods; he differentiates the two concepts with a new framework, where key requirements of design for autonomous systems include the capabilities such as agent viability in a target context, agent self-governance in goal formulation and fulfilment of roles, and independence in defined tasks performance. Kaber (2018) auges that the traditional automation taxonomies (e.g., Sheridan & Verplank, 1978; Parasuraman, Sheridan & Wickens, 2000) were not referring to autonomous agents capable of formulating objectives and determining courses of action and make no mention of, for example, self-sufficiency or self-governance in context.

We argue that non-AI based automated systems and AI-based autonomous systems are not differentiated in terms of level of automation; the essential difference between the two depends on whether or not there is a certain level of human-like cognitive/intelligent capabilities on AI technology. However, AI technology is a double-edged sword. On the one side, with the help of AI technologies (e.g., algorithms, deep machine learning, big data), an AI system can complete certain tasks that cannot be done by previous automation technology in certain scenarios (Kaber, 2018; Madni & Madni, 2018; Xu. 2021); on the other hand, the unique autonomous characteristics of AI systems, such as self-evolving machine behaviors and potential non-deterministic/unexpected outcomes, may cause biased and unexpected system output without human supervision, harming humans and society (Kaber, 2018; Rahwan et al., 2019; Xu, 2021).

### 1.3 An emerging form of the human-machine relationship in the AI era

AI technology has also brought in a paradigmatic change in the human-machine relationship. The focus of HCI work has traditionally been on the interaction between humans and non-AI computing systems. These systems rely on fixed logic rules and algorithms to respond to the input from humans, humans interact with these systems in a form of "stimulus-response" (Farooq, U., & Grudin, J. (2016). These computing systems (e.g., automated machines) primarily work as an assistive tool, supporting human's monitoring and execution of tasks (Wickens et al., 2015).

As we are currently entering in the AI-based "autonomous world," AI technology has given new roles to machines in human-machine systems as driven by its autonomous characteristics. The interaction between humans and AI systems is essentially the interaction between humans and the autonomous/AI agents in the AI systems. As AI technology advances, intelligent agents can be developed to exhibit unique behavior and possess the autonomous characteristics with certain levels of human-like intelligent abilities as summarized in Table 1 (Rahwan et al., 2019; Xu, 2021). With AI technology a machine can evolve from an assistive tool that primarily supports human operations to a potential collaborative teammate of a team with a human operator, playing the dual roles of "assistive tool + collaborative teammate" (e.g., Brill et al., 2018; Lyons et al., 2018; O'Neill et al., 2020). Thus, the human-machine relationship in the AI era has added a new type of human-AI collaboration, often called "Human-Machine Teaming" (e.g., Brill et al., 2018; Brandt et al., 2018; Shively et al., 2018).

The exploratory research work related to human-AI collaboration has begun in some work domains. Examples include intelligent air traffic management system (Kistan et al., 2018), operator-intelligent robot teaming (Calhoun et al., 2018), pilot-system teaming on the intelligent flight deck (Brandt et al., 2018), driver-system teaming in high-level autonomous vehicles (de Visser et al., 2018; Navarro, 2018). We will discuss these in detail in Section 3.2.

Human-machine teaming also has a "double-edged sword" effect. For example, on the one hand, AI technologies (e.g., deep machine learning, big data of collective domain knowledge) can help human decision-making operations be more effective way under some operating scenarios, than individual operators using non-AI systems; on the other hand, if the human-centered approach is not followed in the development of AI systems, there is no guarantee that humans have the final decision-making authority of the systems in unexpected scenarios, and the potential unexpected and indeterministic outcome of the systems may cause ethical and safety failures (Yampolskiy, 2019; McGregor et al., 2021). Thus, AI technology has brought in new challenges and opportunities for HCI design.

**1.4 The research questions and the aims of the paper**

Based on the previous discussions, as AI technology becomes increasingly ubiquitous in people's work and lives, the machines being used by humans are no longer the "conventional" computing systems as we have known them. From the work focus perspective as HCI professionals, we are witnessing the transition from "conventional" human-computer interaction to human interaction with AI-based systems that exhibit new qualities (Rahwan et al., 2019; Kaber, 2018; Xu, 2021). *While the HCI community has contributed to the development of AI systems, what else does it need to do to enable the implementation of the HCAI approach and offer HCI solutions to address the unique challenges posted to humans and society by AI technology?*

This vision of work further challenges the HCI community to prepare for a future that requires the design of AI systems whose behaviors we cannot fully anticipate and for work that we do not know enough about (Stephanidis, Salvendy et al., 2019; Hancock, 2020; Lau et al., 2018). There are three research questions that we intend to answer in this paper:

1. What are the challenges to HCI professionals to develop human-centered AI systems?

2. What are the opportunities for HCI professionals to lead in applying HCAI to address the challenges?

3. How can current HCI approaches be improved to apply HCAI?

The purposes of this paper are to identify the new challenges and opportunities for HCI professionals as we deal with human interaction with AI systems; and to further advance HCAI by urging HCI professionals to take actions addressing the new challenges in their work.

To answer the three questions, we conducted a high-level literature review and analysis, in conjunction with our previous work in promoting HCAI. The rest of this paper is organized as follows: (1) highlight of the challenges for HCI professionals to implement HCAI as we transition to human interaction with AI systems (Section 2); (2) the opportunities for HCAI professionals to address the challenges (Section 3); (4) analyses of the gaps in current HCI approaches for applying

HCAI and our recommendations for closing these gaps (Section 4). We present our conclusions with Section 5.

## 2. New challenges for HCI professionals to develop human-centered AI systems

### 2.1 Identifying new challenges in human interaction with AI systems

Research and applications of human interaction with AI systems are not new; many different research agendas, have been investigated over the past several years. People promoted their work under a variety of labels, such as, human-AI/machine teaming (e.g., Brill et al., 2018; Brandt et al., 2018), human-AI interaction (e.g., Amershi et al., 2019; Yang, Steinfeld et al., 2020), human-agent interaction (e.g., Prada & Paiva, 2014), human-autonomous system interaction (e.g., Cummings & Clare, 2015), human-AI symbiosis (e.g., Nagao, 2019). While there are different focuses across these studies, all of them essentially investigate human interaction with the "machines" in AI systems (i.e., intelligent agents, AI agents, autonomous systems) driven by AI technology; that is, humans interact with AI systems.

Thus, our goal was to answer our first research question: What are the new challenges for HCI professionals to develop human-centered AI systems as we transition from human interaction with conventional non -AI systems to interaction with AI systems? To this end, our high-level literature review and analysis was focused on the following two aspects.

- The research and application work that have been done in human interaction with AI systems

- The unique issues related to AI systems that HCI professionals did not encounter in conventional HCI work (i.e., human interaction with non-AI systems)

The following four electronic databases were used to find the related papers over the last 10 years as of May 2, 2021: American Computing Machinery (ACM) Digital Library, IEEE Xplore Digital Library, Google Scholar, and ResearchGate. As a result, we found about 890 related papers; we categorized the issues covered in these papers into 10 groups, then further categorized the issues into seven groups based on a further analysis with the primary references cited in this paper. We believe that the categorized seven main issues (Table 2) can represent the primary HCI challenges that reveal the significant differences between the familiar HCI concerns of human interaction with non-AI systems (see the Familiar HCI Concerns with Non-AI Systems column in Table 2) and the new HCI challenges of human interaction with AI systems (see the New HCI Challenges column in Table 2).

**Table 2  Summary of main issues for human interaction with AI systems**

| Main Issues | Familiar HCI Concerns with Non-AI Systems (e.g., Jacko, 2012) | New HCI Challenges with AI Systems (Selected references) | Primary HCAI Design Goals (Figure 1) | Detailed Analysis & References (Section#) |
|---|---|---|---|---|
| Machine behavior | • Machines behave as expected by design<br>• HCI design focuses on usability of system output/UI, user mental model, user training, operation procedure, etc. | • AI systems can be developed to exhibit unique machine behaviors with potentially biased and unexpected outcomes. The machine behavior may evolve as the machine learns (Rahwan et al., 2019) | • Human controlled AI | Section 3.1 |
| Human-machine collaboration | • Human interaction with non-AI computing system<br>• Machine primarily works as an assistive tool<br>• No collaboration between humans and machines<br>. | • The intelligent agents of AI systems may be developed to work as teammates with humans to form human-AI collaborative relationships but there is debate on the topic (Brill et al., 2018; O'Neill et al., 2020) | • Human-driven decision-making<br>• Human controlled AI | Section 3.2 |
| Machine intelligence | • By definition, non-AI systems do not have machine intelligence | • With AI technology, machines can be built to have certain levels of human-like intelligence (Watson & Scheidt, 2005)<br>• Machines cannot completely emulate advanced human | • Augmenting human<br>• Human-controlled AI | Section 3.3 |
| | | • cognitive capabilities, developing machine intelligence in isolation encounters challenges (Zheng et al., 2017)<br>• How to integrate human's role into AI systems to ensure human-controlled AI (Zanzotto, 2019) | | |
| Explainability of machine output | • Machine output is typically explainable if the user interface is usable through HCI design | • AI systems may exhibit a "black box" effect that causes the output obscure to users, users may not know how and why AI systems make decisions, when to trust AI (Muelle et al., 2019) | • Explainable AI | Section 3.4 |
| Autonomous characteristics of machines | • Non-AI systems (e.g., automated systems) do not have autonomous characteristics<br>• HCI design focuses on system UI, automation awareness, human-in-loop design, human intervention in emergency | • AI systems may be developed to have unique autonomous characteristics (e.g., learning, self-adaption, self-execution) (Kaber, 2018)<br>• AI systems may handle some operating situations not fully anticipated (O'Neill et al., 2020)<br>• The output of autonomous systems may not be deterministic (Kaber, 2018; Xu, 2021) | • Human-controlled AI | Section 3.5 |
| User interface | • Usability design of cconventional user interface (graphical user interface, visible interface, etc.) | • Intelligent user interface (e.g., voice input, facial / intention recognition)<br>• UI may be invisible & implicit (Streitz, 2007)<br>• How to design intelligent UI usable and natural<br>• AI technology adapts to humans capabilities vs. humans adapt to AI (Cooke, 2018)<br>• The need of HCI design standards specifically developed for AI systems (e.g., Google PAIR, 2019) | • Usable AI | Section 3.6 |
| Ethical design | • Primary user needs include usability, functionality, security | • Ethical issues become more significant, including issues such as privacy, ethics, fairness, skill growth, decision-making authority (e.g., IEEE, 2019). | • Ethical & responsible AI | Section 3.7 |

As listed in Table 2, HCI professionals did not encounter these new HCI challenges in human interaction with non-AI systems. Furthermore, based on the HCAI approach (Xu, 2019a), we specified the primary HCAI-driven design goals for each of the main issues as shown in the Primary HCAI Design Goals column (to be further discussed in Section 2.2). The Detailed Analyses & References column indicates the section number in this paper where we specifically discuss the findings.

It should be noted that the literature review was done in a qualitative nature, it may not be complete and there may be other new challenges yet to be revealed. However, we argue that these findings are sufficient to reveal the trending new challenges and opportunities for HCI professionals when working on human interaction with AI systems, and the findings are sufficient to compel us to make a call to urge HCI professionals to take action addressing the unique issues in AI systems. Also, we assessed these new challenges and opportunities for HCI professionals from a holistic perspective, instead of having a deep dive on individual issues. Such a holistic perspective allows us to develop a strategic view for the future HCI work in the development of AI systems, so that we can offer our strategic recommendations to the HCI community.

Based on the analyses so far, we can draw the following initial conclusions. As AI technology becomes increasingly ubiquitous in people's work and lives, the machine being daily used by humans is no longer just the "conventional" computing system, but AI systems that exhibit unique characteristics post new challenges to HCI design. As HCI professionals, we need to drive human-centered solutions to timely and effectively address these new challenges. The HCI community must understand these new challenges and reassess the scope and methods of current HCI approach on how we can effectively develop human-centered AI systems. Such changes will inevitably bring about a paradigmatic shift for HCI research and application in the AI era, driving new design thinking and HCI approaches in developing AI systems.

## 2.2 Advancing the human-centered AI (HCAI) approach to address the new challenges

While the technology-centered approach is dominating the development of AI technology, researchers have individually explored a range of human-centered approaches to address the unique issues introduced by AI technology. For examples, these include humanistic design research (Auernhammer, 2020), participatory design (Neuhauser & Kreps, 2011), inclusive design (Spencer et al., 2018), interaction design (Winograd, 1996), human-centered computing (Brezillon, 2003; Ford et al., 2015; Hoffman et al., 2004)**,** and social responsibility (Riedl, 2019). Each approach provides a specific perspective and allows examination of specific aspects of an AI system design.

In response to the possible negative ethical and moral impacts of AI systems, in 2018 Stanford University established a "human-centered AI" research institution. The strategy emphasizes on two aspects of work: technology and ethical design. They have realized that the next frontier for the development of AI systems cannot be just technology, it must also be ethical and beneficial to humans; AI is to augment human capabilities rather than replace them (Li, 2018; Donahoe, 2018).

Shneiderman and Xu have moved forward further proposing a human-centered AI (HCAI) approach (Shneiderman, 2020a, 2020b, 2020c; Xu, 2019a, 2019b).  Specifically, Xu (2019a) proposed a comprehensive approach for developing human-centered AI (HCAI). The HCAI approach includes three primary aspects of work interdependently: technology, human factors, and ethics, promoting the concept that we must keep human at the central position when developing AI systems. Shneiderman (2020a) promotes HCAI by providing a two-dimensional design guidelines framework and a comprehensive framework for the governance of ethical AI (Shneiderman, 2020d). Other researchers have also promoted HCAI approach from a variety of perspectives, such as design process (Cerejo, 2021), human-centered explainable AI (Ehsan et al., 2020; Bond et al., 2019), trustworthy AI (He et al., 2021), human-centered machine learning (Sperrle et al., 2021; Kaluarachchi et al., 2021).

However, there are gaps in previous HCAI work. First, most of the previous HCAI work was aimed primarily at either a broad audience or the AI community but not specifically for the HCI community, resulting in uncertainty in what action should be taken by HCI professionals to apply HCAI in their work addressing the new challenges that we did not encounter in human interaction with non-AI systems. Secondly, most of the previous work promoted HCAI at a high level, with no specific HCAI-driven design goals that can be closely tied to the new HCI challenges in human interaction with AI systems. Consequently, HCI professionals could not get direct guidance to their HCI work. Thirdly, there was no holistic assessment that specifically identified the unique challenges as we transition to interaction with AI systems from an HCI perspective, with the result that there are no specific opportunities tied to HCAI-driven design goals identified to guide HCI professionals addressing the new issues in developing HCAI systems. Lastly, the previous HCAI work has not included a comprehensive assessment of current HCI approaches (e.g., methods, process, skillset) on whether we have gaps to enable HCAI in future HCI work.

Thus, to further advance HCAI specifically to the HCI community, we need to specify HCAI-driven design goals, tie these to the unique challenges and opportunities for HCI professionals, and then carry out an assessment of current HCI approaches. We believe that such a goal-driven approach will help further advance the HCAI approach that can guide HCI professionals to take advantage of specific opportunities to address the new challenges in AI systems, eventually developing true HCAI systems.

To achieve the goal of further advancing HCAI, we first specified the HCAI-driven design goals and then tied these goals to the new challenges identified, as shown in the Primary HCAI Design Goals column in Table 2, so that HCI professionals can fully understand what HCAI design goals to aim for when addressing these new HCI challenges. We further elaborate the HCAI framework (Xu, 2019) by mapping these HCAI design goals across the three primary aspects of work interdependently: Technology, Human Factors, and Ethics (see the three smaller circles around the big "Human" circle at the center of Figure 1).

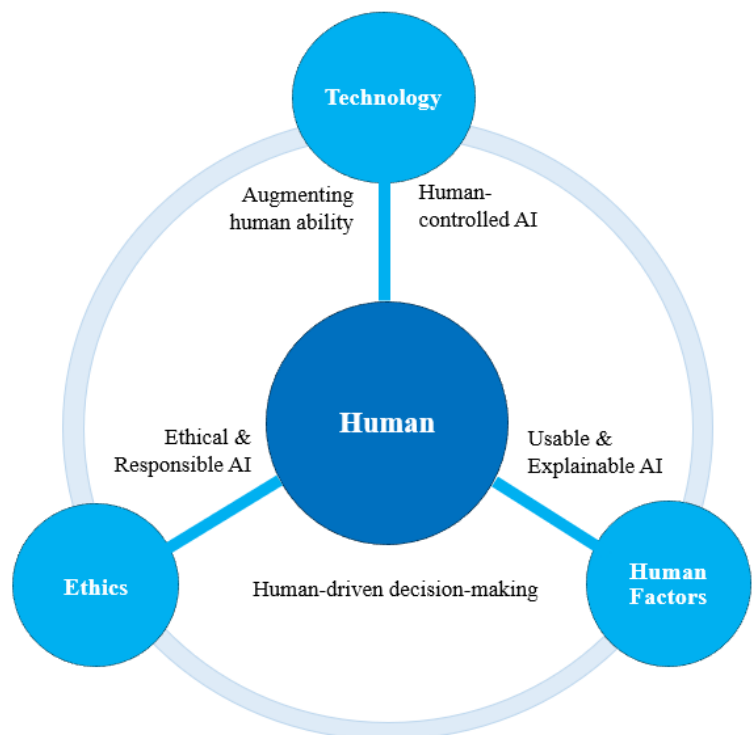

**Figure 1 The Human-Centered AI (HCAI) framework with specified design goals (Adapted from Xu, 2019a)**

(1) "Human Factors": We start from the needs of humans and implement the human-centered design approach advocated by the HCI community (e.g., user research, modeling, iterative user interface prototyping and testing) in the research and development of AI systems. The design goals are to develop usable & explainable AI and AI systems that guarantee human-driven decision-making (see the design goals next to the "Human Factors" circle as depicted in Figure 1).

(2) "Technology": We promote an approach for the development of AI technology by integrating human roles into human-machine systems and by taking complementary advantages of machine intelligence and human intelligence (see 3.3 for details). The design goals are to develop human-controlled AI and augment human abilities, rather than replacing humans (see the design goals next to the "Technology" circle in Figure 1).

(3) "Ethics": We must guarantee fairness, justice, privacy, and accountability in developing AI systems as ethical issues are much more significant in AI systems than conventional non-AI systems. The design goals are to develop ethical & responsible AI and AI systems that guarantee human-driven decision-making (see the design goals next to the "Ethics" circle in Figure 1).

Furthermore, the HCAI approach illustrated in Figure 1 is characterized as follows.

(1) *Placing humans at the center.* All the three primary aspects of work defined in the HCAI approach place humans at the center. Specifically, the "technology' work ensures that humans are kept at the center of AI systems through a deep integration between machine intelligence I and human intelligence to augment human capabilities. The "human factors" work ensures that AI systems are human-controllable by keeping humans as the ultimate decision makers, usable by providing effective interaction design, and explainable by providing understandable output to humans. The "ethics" work aims to address the specific ethical issues in AI systems by delivering ethical AI and providing "meaningful human control" for responsibility and accountability (e.g., van Diggelen et al., 2020; Beckers et al., 2019).

(2) *The interdependency of human factors, technology, and ethics.* The HCAI approach advocates synergy across the three aspects of work as illustrated by the lines connecting all the three aspects depicted in Figure 1. For example, if the impact of AI on humans (e.g., ethics) is not considered in design, it is impossible for AI systems to achieve the human-centered goals and may eventually harm humans even though the AI technology deployed is technically made "more like humans." On the other hand, ethically designed AI systems emphasize augmenting human capabilities rather than replacing humans. AI systems, such as autonomous weapons and vehicles, need to provide an effective meaningful human control mechanism (e.g., human-controllable interface through HCI design) to ensure that the human operators can quickly and effectively take over the control of systems in an emergency (van Diggelen et al., 2020). Also, the system should be able to track the accountability of human operators through the meaningful human control mechanism (see more in Section 3.7).

(3) *Systematic design thinking*. From a technical point of view, the HCAI approach considers humans and machines as a system and seeks to develop the complementarity of machine intelligence and human intelligence within the framework of human-machine systems. From the HCI perspective, the HCAI approach emphasizes on the perspective of human-machine-environment system. An effective AI system should be the optimal match between human needs (physiology, psychology, etc.), AI technology, and environment (physics, culture, organization, society, etc.). From the perspective of ethical design, the HCAI approach systematically considers factors such as ethics, morality, law, and fairness. Therefore, the HCAI approach emphasizes that the development of AI systems is not just a technological project, but a sociotechnical project that needs collaboration across disciplines.

It should be pointed out that we specify the HCAI design goals herein as a minimum, HCI professionals need to collaborate with AI professionals to develop AI systems that meet these specific design goals across the three aspects of their work, then we can achieve the overall goal of HCAI: reliable, safe, and trustworthy AI (Shneiderman, 2020a; Xu, 2019a). The HCAI framework is scalable, implying that we may add more specific design goals across the three aspects of work (i.e., human factors, technology, ethics) to address additional challenges in AI systems as we move forward in the future.

To sum up, Section 2 has answered the first research question: What are the challenges to HCI professionals to develop human-centered AI systems? Based on our high-level literature review and analysis and driven by the HCAI approach, we have identified the unique challenges that HCI professionals need to address in the AI era where humans interact with AI systems across the seven main issues.

The emergence of these AI related unique issues is inevitable as technology advances, just like the emergence of HCI in the 1980's, when HCI professionals faced the issues arise from personal computer technology. Today, we are just facing a new type of "machines"- AI systems that present unique characteristics and challenges to us, urging us to take action. HCI professionals must fully understand the new challenges in human interaction with AI systems, then we can take new design thinking on how to effectively address the new challenges in the development of AI systems.

## 3. New Opportunities for HCI professionals to enable HCAI

The HCI community has recognized that new technologies present both challenges and opportunities to HCI professionals (e.g., Stephanidis, Salvendy et al., 2019; Shneiderman et al., 2016). Specifically, Stephanidis, Salvendy et al. (2019) identified grand challenges and offered comprehensive recommendations on the research priorities to address these grand challenges with intelligent interactive systems. As discussed previously, our approach is to focus on a holistic understanding of the unique challenges of human interaction with AI systems as compared to human interaction with non-AI systems from an HCAI perspective. More specifically, we identify the new HCI opportunities for HCI professionals to address the new challenges based on the HCAI approach and specific HCAI design goals.

This section is organized as follows: (1) highlight of the new challenges across the seven main issues as listed in Table 2 for HCI professionals to address; (2) review of the overall status of current research and application in each of the seven main issues from the HCI perspective; (3) highlight of the new HCI opportunities that HCI professionals can make (or continue to make) contributions to develop AI systems, as driven by the specific HCAI design goals.

### 3.1 From expected machine behavior to potentially unexpected behavior

The machine behavior (system output) of non-AI computing systems is typically expected since design of such systems are based on fixed rules or algorithms. Examples include: washing machines, elevators, and conventional office software. However, the behavioral outcome of AI systems could be non-deterministic and unexpected. Researchers are raising the alarm about the unintended consequences, which can produce negative societal effects (e.g., Yampolskiy, 2019; McGregor et al., 2021). Machine behavior in AI systems also has a special ecological form (Rahwan et al., 2019). Currently, the majority of people studying machine behavior are AI and computer science professionals without formal training in behavioral science (Epstein et al., 2018; Leibo et al., 2018). The leading researchers in the AI community call for other disciplines to join (Rahwan et al., 2019).

The HCAI design goal is to develop human-controlled AI through manageable machine behavior in AI systems in order to avoid biased and unexpected machine behaviors (see Figure 1). Obviously, the multi-disciplinary HCI can play a significant role here (Bromham et al., 2016). HCI professionals can help ensure that we avoid generating extreme or unexpected behaviors and reduce biases generated from AI technologies such as machine learning (Zhang et al., 2020; Leibo et al., 2018). This is also aligned with the "ethical AI" design goal defined in the HCAI approach (Figure 1). Machine behavior is one of the unique characteristics that distinguishes the HCI work from the conventional HCI work for non-AI systems. HCI professionals must fully understand the design implications of machine behavior in developing AI systems and find effective ways to manage the unique behaviors of AI systems in order to deliver HCAI-based systems. Driven by the HCAI approach, HCI professionals need to put humans and their needs first in the development process (e.g., data collection, algorithm data training, test) and to partner with AI professionals to manage the development of machine behavior.

With many of today's AI systems being derived from machine learning methods, study of the mechanism behind a machine's behavior will require assessment of how machine learning methods and processes impact machine behavior (Ribeiro et al., 2016). Many AI systems are based on supervised, reinforced, and interactive learning approaches that require work done by humans, such as, providing an enormous amount of labeled data for algorithm training. Initial work has started. As an example of HCAI-aligned design approaches, based on a "human-centered machine learning" approach (Fiebrink et al., 2018; Kaluarachchi et al., 2021), Interactive Machine Learning aims to avoid concerns over fairness, accountability, and transparency (Fiebrink et al., 2018). The interactive machine learning approach allows users to participate in the algorithm training iteratively by selecting, marking, and/or generating training examples to interact with required functions of the system, supported by HCI design and evaluation methods. The interactive machine learning approach also pays special attention to the interaction between humans and the machine learning process. From the HCAI perspective, the interactive machine learning approach emphasizes on human's roles in the development with human goals and capabilities, so that an AI system can deliver better results than humans or algorithms working alone (Amershi et al., 2014).

The unique machine behavior creates challenges for developing AI systems, but also opportunities for HCI professionals to play a role in future HCI work.

*Application of HCI approach for managing machine behavior*. HCI professionals may leverage HCI methods (e.g., iterative design and evaluation with end users) to continuously improve the design until undesirable results are minimized. A machine acquires its behavior during the development with AI technology; shaping the behavior is affected by many factors, like algorithms, architecture (e.g., learning parameter, knowledge representation), training, and data. For instance, HCI professionals may translate user needs into data needs and understand how to generate, train, optimize, and test AI-generated behaviors during development. Machine behavior can be shaped by exposing it to particular training data. Substantial human effort is necessary to annotate or characterize the data in the process of developing autonomous capabilities (e.g., labeling that a pedestrian is in the scene) (Acuna et al., 2018; Amershi et al., 2014). For instance, classification algorithms for images and text are trained to optimize accuracy on a specific set of human-labeled datasets, the choice/labeling of the dataset and those features represented can substantially influence the behavior (Bolukbasi et al., 2016; Buolamwini et al., 2018). HCI professionals need to figure out an effective approach that can bring in users' expectations to tune the algorithm for preventing biased responses.

*Continued improvement of AI systems during behavior evolution*. HCI professionals needs to study the evolution of the AI-based machine behavior for continued improvement of AI systems. For instance, after being released to the market, algorithms, learning, and training data will all affect the evolution of an AI systems' behaviors. An autonomous vehicle will change its behaviors over time by software and hardware upgrades; a product recommendation algorithm makes recommendations based on continuous input from users, updating their recommendations accordingly. The AI community recognizes that it is not easy to define the expected result due to user's subjective expectations (Pásztor, 2018), HCI professionals need to find a collaborative method on how to improve the design through re-training based on user feedback and how to collect training data and define the expected results that users hope to obtain from AI systems, continuously improving the design of AI systems.

*Enhancement of software testing approach*. The testing methods in traditional software engineering are based on predictable software output and machine behaviors, which was specifically created for testing non-AI computing systems. Since the outputs of AI systems will evolve as the systems learn over time, HCI professionals need to collaborate with AI and computer science professionals to find ways to improve the testing methods from a behavioral science perspective. Due to the existence of the unique machine behavior, how do we effectively test and measure the evolving performance of AI systems by taking humans and AI agents as a human-machine system? As AI technology advances, it is also critical to assess its influence on humans and society on a long-term basis (e.g., a longitudinal perspective), requiring enhanced testing methods to be supported by social scientist and psychologists (Horvitz, 2017; Stephanidis, Salvendy et al., 2019).

*Leveraging HCI approach in design*. HCI professionals can partner with AI professionals to develop better AI systems from an HCI design perspective. For instance, for the interactive machine learning approach, the system displays the current model and corresponding results (such as classification results, clustering results, prediction results, etc.) to users through the user interfaces of AI systems, HCI professionals can support the design from the perspectives of visualization design, interactive technology and mental modeling of target users. Also, many human-centered machine learning research projects conducted by AI professionals made AI engineers the 'human' at the center in their design, instead of end users (Kaluarachchi et al., 2021), HCI professionals need to ensure target end users to be fully considered based on the HCAI approach.

## 3.2 From interaction to potential human-AI collaboration

As discussed earlier, there is an emerging form of the human-machine collaboration in the AI era: human- machine teaming. As HCI professionals, we are not just dealing with the conventional the "interaction" between humans and machines, but also a new form of a human-machine relationship that requires new perspective to study.

A significant amount of work has been invested in the area of human-machine teaming. Initial research suggests that humans and AI systems can be more effective when working together as a combined unit rather than as individual entities (Bansal et al., 2019b, 2019b; Demir et al., 2018b). People argue that the more intelligent the AI system, the greater the need for collaborative capabilities (Johnson & Vera, 2019). Designing for such potential collaboration suggests a radical shift from the current HCI thinking to a more sophisticated strategy based on teaming (Johnson & Vera, 2019).

A variety of topics are being explored in the research and application of the potential human-AI collaboration, such as, conceptual architecture and framework (e.g., Madni & Madni, 2018; Johnson & Vera, 2019; O'Neill et al., 2020; Prada & Paiva, 2014), performance measurements (e.g., Bansal et al., 2019a). To study the human-AI collaboration, researchers have leveraged the frameworks of other disciplines, such as psychological human-human team theory (e.g., de Visser, 2018; Mou & Xu, 2017). For example, the human-human team theory helps formulate basic principles: two-way/shared communication, trust, goals, situation awareness, language, intentions, and decision-making between humans and AI systems (e.g., Shively et al., 2018; Ho et al., 2017; Demir et al., 2017), instead of a one-way approach as we currently do in the conventional HCI context.

There are debates on whether AI systems can truly work as a teammate with humans (e.g., Schneiderman, 2021c; Klein, G., Woods, et al., 2004), we believe that a common HCAI design goal should be shared between both sides to ensure human-controlled AI and human-driven decision-making in human interaction (or collaboration) with AI (see Figure 1). Humans should not be required to adapt to non-human "teammates"; instead, designers should create technology to serve as a good team player (or super tool) alongside humans (Cooke, 2018; Schneiderman, 2021c). HCI professionals need to fully understand what the machines really can do. There is a need for future HCI work in this area.

*Clarification of human and AI's roles*. While HCI traditionally studied the interaction between humans and non-AI computing systems, future HCI work should be focused on understanding of how humans and AI systems interact, negotiate, and even work as teammates for collaboration. Future HCI work needs to assess whether humans and AI agents can be a true collaborative teammate versus an AI agent serving merely as a super tool (Shneiderman, 2020c; Klein, G., Woods, et al., 2004), as a peer (Ötting, 2020), or as a leader (Wesche & Sonderegger, 2019). As AI technology advances, an important HCI question is the design-center for the interaction (or mutual collaboration); that is, who are the ultimate decision makers? We need to investigate how we can ensure that when in complex or life-changing decision-making, humans retain a critical decision-making role instead of AI systems while creating and maintaining an enjoyable user experience (Inkpen et al., 2019).

*Modeling human-AI interaction and collaboration*. Conventional interaction models (e.g., MHP, GOMS) can no longer meet the needs of complex interactions in the AI era (e.g., Card et al., 1983), there is a need for HCI professionals to explore the human cognitive mechanisms for modeling the interaction and collaboration with AI systems (Liu et al., 2018). The goal is to design interfaces of AI systems for facilitating effective interaction and potential collaboration. Other topics include the models of teaming relationships (e.g., peer, teammate, and mentor/mentee), teaming processes (e.g., interactions, communications, and coordination), and applications of human-AI collaboration across different work domains.

*Advancement of current HCI approach*. HCI professionals should not only rely on current HCI approach but do much more. For example, Computer-Supported Cooperative Work (CSCW) has long been an active subfield of HCI, and its goal is to use the computer as a facilitator to mediate communication between people (Kies et al., 1998). Lieberman (2009) argues that AI brings to the picture of a collaboration between the user and the computer taking a more active collaborative role, rather than simply a conduit for collaboration between people. We need to do a comparative assessment between CSCW and the potential human-AI collaboration, and fully understand the implications for system design.

*Innovative HCI design to facilitate human-AI collaboration*. Future HCI work also needs to explore innovative design approaches (e.g., collaborative interaction, adaptive control strategy, human directed authority) from the perspective of human-AI interaction. For instance, we need to model under what conditions (e.g., based on a threshold of mutual trust between humans and AI agents) an AI agent will take over or hand off the control of system to a human in specific domains such as autonomous vehicles (Kistan et al., 2018; Shively et al., 2017). In the context of distributed AI and multi-agent systems, we need to figure out the collaborative relationship between at least one human operator and the collective system of agents, where and how multiple agents communicate and interact with primitives such as common goals, shared beliefs, joint intentions, and joint commitments, as well as conventions to manage any changes to plans and actions (Wooldridge, 2009; Hurts et al., 1994).

### 3.3 From siloed machine intelligence to human-controlled hybrid intelligence

With the development of AI technology, the AI community has begun to realize that machines with any degree of intelligence cannot completely replace human intelligence (e.g., intuition, consciousness, reasoning, abstract thinking), and the path of developing AI technology in isolation has encountered challenges. Consequently, it is necessary to introduce the role of human or human cognitive model into the AI system to form a hybrid intelligence (e.g., Zheng et al., 2017).

Researchers have explored hybrid intelligence by leveraging the strengths from both human intelligence and machine intelligence (Zheng et al., 2017; Dellermann et al., 2019a, 2019b; Johnson & Vera, 2019). Recent work has initially found the significant potential of augmentation through integration of both kinds of intelligence. The applications of hybrid augmented intelligence help achieve collaborative decision-making in complex problems, thereby gaining superior excellent results that cannot be achieved separately (e.g., Carter & Nielsen, 2017; Crandall et al., 2018; Dellermann, et al., 2019a, 2019b).

The HCAI design goal is to develop human-controlled AI through human-machine hybrid intelligence (see Figure 1). Driven by the HCAI approach, we advocate that hybrid intelligence must be developed in a context of "human-machine" systems by leveraging the complementary advantages of AI and human intelligence to produce a more powerful intelligence form: human-machine hybrid intelligence (Dellermann et al., 2019b). This strategy not only solves the bottleneck effect of developing AI technology as discussed earlier, but also emphasizes the use of humans and machines as a system (human-machine system), introducing human functions and roles, and ensuring the final decision of humans on the system control.

In the AI community, the research on hybrid intelligence can basically be divided into two categories. The first category is to develop human-in-the-loop AI systems at the system level, so that humans are always kept as a part of an AI system (e.g., Zanzotto, 2019). For instance, when the confidence of the system output is low, humans can intervene by adjusting input, creating a feedback loop to improve the system's performance (Zheng et al., 2017; Zanzotto, 2019). Human-in-the-loop AI systems combines the advantages of human intelligence and AI, and effectively realizes human-AI interactions through user interventions, such as online assessment for crowdsourced human input (Mnih et al., 2015; Dellermann et al., 2019a), user participation in training, adjustment and testing of algorithms (Acuna et al., 2018; Correia et al., 2019).

The second type of hybrid intelligence involves embedding the human cognitive model into an AI system to form a hybrid intelligence based on cognitive computing (Zheng et al., 2017). This type of AI system introduces a variety of biologically inspired computing intelligence technologies. Elements include intuitive reasoning, causal models, memory and knowledge evolution, etc. According to the HCAI approach, this type of hybrid intelligence is not in the true sense of human-machine hybrid intelligence, because such systems are not based on human-machine systems and cannot be effective in achieving the central decision-making role of human operators in the system. Of course, the introduction of human cognitive models into machine intelligence, is important for the development of machine intelligence.

There will be a lot of opportunities for HCI professionals to play a role in developing human-machine hybrid intelligence systems.

*Human-controlled hybrid intelligence*. The HCAI approach advocates human-in-the loop AI systems, but we emphasize that the systems must be based on human-controlled AI. In other words, AI systems are embedded in the "human loop" as supporting super tools (Shneiderman, 2021c), ensuring that human-driven decision-making is still primary. It has been argued that there are currently two basic schemes in terms of the control mechanism (who is in control) of hybrid intelligence: "human-in-loop control" and "human-machine collaborative control" (Wang, 2019). The effective handover between humans and AI systems in an emergency state is a current important topic. The loss of control in autonomous systems (e.g., decision tracking after launch of autonomous weapon systems, autonomous vehicles) has become a widely concerned issue from the perspective of the "ethical AI" design goal defined in the HCAI approach. The HCAI approach requires humans to have the ultimate control. HCI needs to carry out this work from the HCI perspectives. Future HCI work needs to seek solutions from an integrated perspectives of system design, human-machine interaction, and ethical AI design.

*Cognitive computing and modeling*. As an interdisciplinary, future HCI work needs to help the AI community explore cognitive computing based on human cognitive abilities (e.g., intuitive reasoning, knowledge evolution) in support of developing human-machine hybrid intelligent systems. Future HCI work should help accelerate the conversion of existing psychological research results to support the work of cognitive computing and define cognitive architecture for AI research (Salvi et al., 2016). This is essential because the development of existing hybrid intelligent systems based on cognitive computing does not fully consider the central role and decision-making function of human operators in the systems. HCI professionals should collaborate with AI professionals to explore the approach of integrating the cognitive computing method with the human-in-the-loop method, either at both system and/or at biological levels (e.g., brain-computer interface technology), developing HCAI-based hybrid intelligent systems.

*HCI framework for human-machine hybrid intelligence*. Future HCI work needs to lay the foundation for developing effective human-machine hybrid intelligent systems in terms of theory and methods. For example, we may apply the joint cognitive systems theory (Hollnagel & Woods, 2005) and consider a human-machine intelligent hybrid system as a joint cognitive system with two cognitive agents: human and machine. Research work is needed to explore how to use the two cognitive agents to support the design of autonomous systems based on the concept of human-machine hybrid intelligence through a deep integration between human biological intelligence and machine intelligence.

*Human role in future human-computer symbiosis*. Licklider (1960) proposed the well-known concept of human-computer symbiosis, pointing out that the human brain could be tightly coupled with computers to form a collaborative relationship. In the long run, people argue that the human-machine hybrid intelligence will form an effective productive partnership to achieve a true human-machine symbiosis at both systemic and biological levels, eventually enabling a merger between humans and machines (e.g., Gerber et al., 2020). HCI professionals need to ensure that such a human-machine symbiosis will deliver the most value to humans and fully ensure that human functions and roles are seamlessly integrated into AI systems as the ultimate decision makers without harming humans, eventually moving towards a so called "human-machine symbiosis world" with humans at the center (Gerber et al., 2020), instead of a "machine world".

### 3.4 From user interface usability to AI explainability

Designing a usable user interface (UI) for a computing system is the core work for HCI professionals in developing non-AI systems. We currently encounter new challenges in designing the user interfaces for AI systems besides UI usability. As the core technology of AI, machine learning and its learning process are opaque, and the output of AI-based decisions is not intuitive. For many non-technical users, a machine learning-based intelligent system is like a "black box", especially neural networks for pattern recognition in deep machine learning. This "black box" phenomenon may cause users to question the decisions from AI systems: why do you do this, why is this the result, when did you succeed or fail, when can I trust you? The "black box" phenomenon has become one of the major concerns in developing AI systems (Muelle et al., 2019; Hoffman et al., 2018; Bathaee, 2018; Gunning, 2017). The "black box" phenomenon may occur in various AI systems, including the application of AI in financial stocks, medical diagnosis, security testing, legal judgments, and other fields, causing inefficiency in decision-making and loss of public trust in AI.

As a solution, explainable AI intends to provide the user what the machine is thinking and can explain to the user why (PwC, 2018), which is also the HCAI design goal (see Figure 1). Explainable AI is getting more and more attention. One representative research program is the five-year research plan for explainable AI launched by DARPA in 2016 (Gunning, 2017). The program covers three aspects of work: (1) develop a series of new or improved machine learning technologies to obtain explainable algorithms; (2) develop explainable UI model with advanced HCI technology and design strategies; and (3) evaluate existing psychological theories on explanation to assist explainable AI.

Existing psychological theories on explanation may help the explainable AI work. Psychological research has been carried out in the aspects of mechanisms, measurements, and modeling. For example, inductive reasoning, causal reasoning, self-explanation, comparative explanation, counterfactual reasoning (Hoffman, et al., 2017; Broniatowski, 2021). Although many hypotheses were verified in laboratory studies (Mueller, et al., 2019), much work is still needed to build computational models for explainable AI work. Researchers have realized the importance of non-AI disciplines such as HCI in explainable AI research (e.g., Bond et al., 2019; Hoffman et al., 2018). For instance, the survey of Miller et al. (2017) indicates that most explainable AI projects were only conducted within the AI discipline. AI professionals tend to build explainable AI for themselves rather than target end users; most explainable AI publications do not follow a human-centered approach (Kaluarachchi et al., 2021). Miller et al. (2017) argue that if explainable AI adopts appropriate models in behavioral science, and these evaluations focus more on users rather than technology, explainable AI research is more likely to be successful. These findings show the relevance of the HCAI approach and why HCI professionals should participate in explainable AI research.

While research on explainable AI is currently underway, some researchers have realized that the ultimate purpose of explainable AI should ensure that target users can understand the output by AI systems (Kulesza et al. 2015; Broniatowski, 2021). For example, an "explainable AI" version of data scientists is not understandable to most people. Chowdhury (2018) argues that understandable AI requires the collaborations across disciplines besides AI and data science. The solution should meet target user needs (e.g., knowledge level), and ultimately achieving an understandable AI, which is aligned with HCAI. The current work on understandable AI focuses on UI modeling, applications of psychological theories, and validation of related experiments (Zhou et al., 2018; Hoffman et al., 2018; DARPA, 2017).

Future work cannot rely solely on technical methods and requires the participation of HCI. Future HCI work can be mainly considered as follows.

*Effective interaction design*. Most previous explainable AI research was based on static, one-way interpretation from AI systems. HCI research on the interaction design of AI systems needs to explore alternative approaches, such as co-adaptive dialog-based interface, UI modeling, natural interactive dialogue. Future HCI work needs to continue to develop effective human-centered solutions, such as: interactive dialogues with AI systems (Abdul et al., 2018), visualization approach (Chatzimparmpas et al., 2020), human-in-the-loop explainable AI (Chittajallu et al, 2019), collaborative task-driven (e.g., error detection), and exploratory interpretation to support the adaptation between humans and machines (Kulesza's at al., 2015; Mueller et al., 2019).

*Human-centered explainable AI*. Future HCI work need to further explore user-participatory solutions based on the HCAI approach, such as user exploratory and user-participation approaches (Mueller et al., 2019). This is to ensure that AI is understandable to target users beyond explainability, which AI professionals have previously neglected. In addition, there is a lack of user-participated experimental validation in many existing studies, along with a lack of rigorous behavioral science-based methods (Abdul et al., 2018; Mueller et al., 2019). HCI should give full play to its expertise to support validation methods. When adopting user-participated experimental evaluation, it is necessary to overcome the unilateral evaluation methods in some existing studies that only evaluate the performance of AI systems, HCI should promote the evaluation of the AI systems as a human-machine system by adding an end users perspective (Muelle et al., 2019; Hoffman, et al., 2018).

*Acceleration of the transfer of psychological theories*. HCI professionals can leverage our own multidisciplinary backgrounds to act as intermediaries, accelerating the transfer of the knowledge from other disciplines. For example, many of the existing psychological theories and models have not generated usable results (Abdul et al., 2018; Muelle et al., 2019). Also, there is a large body of human factors research work on automation transparency and situation awareness (e.g., Sarter & Woods, 1995; Endsley, 2017), their comprehensive mitigation solutions may offer possible solutions to explainable AI.

*Sociotechnical systems approach.* From a sociotechnical systems perspective, the impacts of other social and organizational factors (e.g., user/organizational trust and acceptance) on explainability and comprehensibility of AI should be further explored (Ehsan & Riedl, 2020; Klein, et al., 2019). HCI professionals need to consider the influence of other factors, such as culture, user knowledge, decision-making behaviors, user skill growth, users' personality and cognitive styles. HCI can apply its multidisciplinary expertise in this regard.

## 3.5 From human-centered automation to human-controlled autonomy

We are transitioning to an "autonomous world". AI technology demonstrates its unique autonomous characteristics, society is currently introducing more autonomous technologies that possess unique characteristics different from traditional automation technologies as discussed earlier, but safety and negative effects have not attracted enough attention (Hancock, 2019; Salmon, 2019). As an example, many companies have heavily invested in developing AI-based autonomous systems (e.g., autonomous vehicles), but consumer's trust on AI is still questionable. Lee and Kolodge's survey (2018) shows that among the 8571 American drivers surveyed, 35 % of them said they definitely do not believe that autonomous vehicles have the ability to operate safely without driver control. On the other side, some consumers over trusted AI-based autonomous systems as evidenced by several fatal accidents of autonomous vehicles (e.g., NTSB, 2017; Endsley, 2018). Researchers have expressed their deep concerns about the safety issues caused by autonomous technologies (e.g., de Visser et al., 2018; Endsley, 2017, 2918; Hancock, 2019).  Also, as we just enter the AI age, people have begun to confuse the concepts of automation and autonomy, leading to inappropriate expectations and potential misuse of technology (Kaber, 2018).

It is also important for HCI professionals to understand how previous research on automation provides lessons and implications to what we should do with autonomous systems. In the past few decades, the human factors community has carried out extensive research on automation and promoted a human-centered automation approach (e.g., Billings, 1997; Sarter & Woods, 1995; Xu, 2007). The results show that users have over-trust and over-reliance on automation and many automation systems have vulnerabilities in complex domains, such as aviation and nuclear power. These systems worked well under normal conditions, but it might cause the human operator's "automation surprise" problems when unexpected events occurred (e.g., Sarter & Woods, 1995; Xu 2004). Operators may not be able to understand what and why automation is doing, creating an "out-of-the-loop" effect and resulting in the wrong manual intervention (Endsley, 2015; Endsley & Kiris, 1995; Wickens & Kessel, 1979). This vulnerability of automation has brought challenges to safety, for example, civil flight deck automation has caused deadly flight accidents (Endsley, 2015). Bainbridge (1983) defined a classic phenomenon of "ironies of automation": the higher the degree of automation, the less the operator's attention to the system; in an emergency state, it is more difficult for the operator to control the system through manual control.

Endsley (2017) believes that in an autonomous system, with the improvement of the "automation" level of individual functions, and as the autonomy of the overall system increases, the operator's attention to these functions and situation awareness will decrease in emergency situations, so that the possibility of the "out-of-the-loop" effect will also increase. Fatal accidents involving autonomous vehicles that have occurred in recent years have showed these typical human factors issues as seen in automation (e.g., Navarro, 2018; NHTSA, 2017). More importantly, AI-based autonomous systems' human-like intelligent abilities (e.g., learning) will continue to evolve as they are used in different environments. The uncertainty of the operational results of autonomous systems means that the system may develop behaviors in unexpected ways. Therefore, initial research shows that autonomous systems may give operators a stronger shock than "automation surprise" (Prada & Paiva, 2014; Shively et al., 2017). These effects may further amplify the impact of the "automation surprise" issue and cause decrements in failure performance and a loss of situation awareness associated with increasing degree of automation, just as suggested by the lumberjack effect (Onnasch et al., 2014). In addition, initial studies also show that autonomous systems may also lead to a highly emotional reaction of the operator, some social factors are more likely to affect the operator's cognitive ability, personality traits and communication attributes (Clare et al., 2015; Mou & Xu, 2017).

From the HCAI perspective, we advocate human-controlled AI through human-controlled autonomy (see Figure 1). At present, we are in the transition from human-centered automation to human-controlled autonomy. The effort to address the classic "ironies of automation" issue has existed over 30 years, but is still not completely solved (Bainbridge, 1983; Strauch, 2017). Today, we encounter new ironies: autonomous systems that exhibit unique characteristics compared with traditional automation. The potentially amplified impacts of autonomy plus its unique autonomous characteristics compels us to take the lessons learned from automation and explore new approaches beyond what we have done in automation to address the new challenges, driving towards human-controlled autonomy and avoiding excessive autonomy (Shneiderman, 2020a), based on the HCAI design goal of human-controlled AI (see Figure 1).

Overall, the HCI research on autonomous systems is currently in its infancy. New challenges and opportunities co-exist as we enter the AI-based autonomous world, requiring HCI professionals go beyond current design thinking of the traditional automation technology and develop innovative approaches to address the unique issues in autonomy brought about AI technology.

*Understanding the impacts of autonomous characteristics.* We need to assess and fully understand the impacts of the autonomous characteristics from the HCAI perspective. There are many basic questions for HCI professionals to explore, for instance, what are the implications for HCI design? We also need to further empirically assess the "automation surprise" effect and the "lumberjack" effect for AI systems with a varied degree of autonomy. The results will inform HCI recommendations to the development of AI-based autonomous systems.

*Innovative design paradigms.* We need to explore alternate paradigms to optimize the design of autonomous systems. Although HCI professionals have participated in the development of autonomous systems (e.g., autonomous vehicles), recent accidents have sounded an alarm, compelling us to rethink the current approach (Xu, 2020). For example, from the human-AI collaboration perspective, the research on human-autonomy teaming has been carried out over the last several years (e.g., Shively et al., 2017; O'Neill et al., 2020). We may explore the design for autonomous systems that emulate the interactions between people by considering mutual trust, shared situation awareness, and shared autonomy/control authority in a reciprocal relationship between human and autonomous systems. Thus, we can maximize design opportunities to minimize the potential risk caused by autonomous systems. Future HCI work may explore applying the paradigm of human-machine hybrid intelligence to develop innovative solutions for effective human-machine co-driving and take-over/handoff (Jeong, 2019). For interaction design, we need to take innovative approaches beyond traditional HCI design. By applying the HCAI approach, we need to develop an effective human control mechanism with well-designed UIs to enable human operators to monitor and quickly take over control of autonomous systems in emergencies.

*Design for human-controlled autonomy*. SAE J3016 (2018) regulations presume that when autonomous vehicles are equipped with high-level (L4-L5) automated driving functions, driver monitoring and manual intervention are not required. HCI professionals need to step in to ensure that humans are always the ultimate decision makers in system design (including remote operations) to address safety issues (Navarro, 2018; Shneiderman, 2021c). To reinforce the "human-controlled AI" design goal, we advocate to implement a "human meaningful control" design to implement the accountability and traceability of autonomous systems (de Sio & den Hoven, 2018). When defining the levels of "automation" for autonomous vehicles, SAE J3016 (2018) seems to be ignoring the differences between automation and AI-based autonomous technology, and SAE J3016 takes a system-centered approach classifying the "automation" level, instead of a human-centered approach (Hancock, 2019). HCI professionals need to obtain empirical data to help improve the classification of autonomous levels from the HCAI perspective, which may help standardize the way of defining product requirements, measurement for certification of autonomous technology, and the need for human operator training.

*A multidisciplinary HCI approach.* As an interdisciplinary area, HCI professionals are in a unique position to address the emerging issues from a broad perspective of sociotechnical systems, including the impacts of autonomous systems on human expectations, operational roles, ethics, privacy, and so on. While we have an established process of certifying humans as autonomous agents, there is no industry consensus as to how to certify computerized autonomous systems for human interaction with AI systems (Cummings, 2019; Cummings & Britton, 2020). More research is needed to further assess the impacts of autonomous technology across a variety of context.

### 3.6 From conventional interactions to intelligent interactions

In the AI era, we are transitioning from conventional interaction with non-AI computing systems to the intelligent interaction with AI systems driven by AI technology, such as voice input and facial recognition. Historically, interaction paradigms have guided UI development in HCI work, e.g., the WIMP (window, icon, menu, pointing) paradigm. However, WIMP's limited sensing channels and unbalanced input/output bandwidth restrict the human-machine interaction (Fan et al., 2018). Researchers explored the concepts of Post-WIMP and Non-WIMP, such as the PIBG paradigm (physical object, icon, button, gesture) for pen interaction (Dai et al., 2004), and RBI (reality-based interaction) for virtual reality (Jacob et al., 2008), but the effectiveness of these proposed paradigms requires verification (Fan et al., 2018). To further analyze the interaction paradigm for AI systems, Fan et al. (2018) proposed a software research framework, including interface paradigms, interaction design principles, and psychological models. Zhang et al. (2018) proposed a RMCP paradigm that includes roles (role), interactive modal (modal), interactive commands (commands), and information presentation style (presentation style) for AI systems. These efforts have not been fully validated and primarily initiated by the computer science community without the participation of HCI professionals. HCI professionals face new challenges, the HCAI design goal of "Usable AI" calls for future HCI work (Figure 1).

*New interaction paradigms*. In the realization of multi-modality, natural UI, and pervasive computing, hardware technology is no longer an obstacle, but the user's interaction capabilities have not been correspondingly supported (Streitz, 2007). How to design effective multi-modal integration of visual, audio, touch, gestures, and parallel interaction paradigms for AI systems is an important topic for future HCI work. This work was mainly carried out in the computer science community, the HCI community should support for defining the paradigms, metaphors, and experimental verification to address the unique issues in AI systems as we did in the personal computer era (Fu et al., 2014; Sundar, 2020).

*Usable user interface.* AI technology changes the way we design user interfaces. There is a big opportunity for HCI professionals to design usable user interface based on effective interaction design with AI systems. Conventional non-AI interfaces tend to be rigid in the order in which they accept instructions and the level of detail which they require (Lieberman, 2009). AI-based technologies, such as gesture recognition, motion recognition, speech recognition, and emotion/intent/context detection, can let systems accept a broader range of human input through multi-modalities in parallel. Human interaction with AI systems, which is distinct from interaction with non-AI systems, requires HCI professionals to develop more effective approaches to support HCI design work exploring the design of usable UI that can effectively facilitate human-AI interaction, including potential human-AI collaboration.

*Adapting AI technology to human capability*. Human limited cognitive resources become a bottleneck of HCI design in the pervasive computing environment. As early as 2002, Carnegie University's Aura project research showed that the most precious resource in pervasive computing is not computer technology, but human cognitive resources (Garlan et al., 2002). In an implicit interaction scenario initiated by ambient intelligent systems, intelligent systems may cause competition between human cognitive resources in different modalities, and users will face a high cognitive workload. In addition, implicit, multi-modal, and pervasive interactions are anticipated to be not only conscious and intentional, but also subconscious and even unintentional or incidental (Bakker & Niemantsverdriet, 2016). Thus, HCI design must consider the "bandwidth" of human cognitive processing and their resource allocation while developing innovative approaches to reduce user cognitive workload through appropriate interaction technology, adapting AI technology to human capabilities as driven by the HCAI approach, instead of adapting humans to AI technology.

*HCI design standards for AI systems*. We also need interaction design standards to guide HCI design work in the development of AI systems. Existing HCI design standards were primarily developed for non-AI systems, there is a lack of design standards and guidelines that specifically support HCAI-based systems. The design standards and guidelines for AI systems need to fully consider the unique characteristics of AI systems, including design for transparency, unpredictability, learning, evolution, and shared control (Holmquist, 2017); design for engagement, decision-making, discovery, and uncertainty (Girardin & Lathia, 2017). There are initial design guidelines available, such as "Google AI + People Guidebook" (Google PAIR, 2019), Microsoft's 18 design guidelines (Amershi et al., 2019), and Shneiderman's 2-D HCAI design framework (Shneiderman, 2020a). The International Organization for Standardization (ISO) also sees this urgency, publishing a document on AI systems, titled "Ergonomics of Human-System Interaction - Part 810: Robotic, Intelligent and Autonomous Systems" (ISO, 2020). The HCI community needs to put forward a series of specific design standards and guidelines based on empirical HCI studies, playing a key role in developing HCI design standards for HCAI systems.

**3.7 From general user needs to specific ethical AI needs**

The focus of the conventional HCI work is on general user needs in interaction with non-AI computing systems, such as usability, functionality, and security. As one of the major challenges presented by AI technology, ethical AI design has received a lot of attentions in research and application, and multiple sets of ethical guidelines from various organizations around the world are currently available, large high-tech companies have published internal ethical guidelines for development of AI systems (e.g., IEEE, 2019; Hagendorff, 2020). Shneiderman (2021d) proposes 15 recommendations at three levels of governance: team, organization, and industry, aiming at increasing the reliability, safety, and trustworthiness of HCAI systems. However, research also shows that the effectiveness of the practical application of ethical AI guidelines is currently far from satisfactory (e.g., Mittelstadt, 2019). In this section, we approach the ethical AI issues from the HCI design perspective with the hope that ethical AI issues can be successfully addressed in the frontline of developing AI system.

Specifically, the concept of "meaningful human control" is one of the approaches being studied (e.g., van Diggelen et al., 2020; Beckers et al., 2019). In alignment with HCAI, van Diggelen et al. (2020) define meaningful human control as having three essential components: (1) human operators are making informed and conscious decisions about the use of autonomous technology; (2) human operators have sufficient information to ensure the lawfulness of the action they are taking; and (3) the system is designed and tested, and human operators are properly trained, to ensure effective control over its use (e.g., autonomous weapons). Currently there are many perspectives for consideration of the meaningful human control, ranging from accountability and system transparency design (Bryson and Winfield, 2017), governance and regulation (Bryson & Theodorou, 2019), delegation (van Diggelen et al., 2020), to certification (Cummings, 2019). There is a consensus that the human must always remain in control of ethically sensitive decisions (van Diggelen et al., 2020; de Sio & den Hoven, 2018), which is also the HCAI design goal (see Figure 1).

The HCI community can help address the ethical AI issues from the following aspects.

*Adoption of the meaningful human control in design*. One of the key requirements for meaningful human control is allowing human operators to make informed and conscious decisions about the use of autonomous technology. To achieve this design goal, HCI professionals need to partner with system designers to ensure a transparent system design and effective interaction design, keeping human operators in the loop with sufficient situation awareness. Also, HCI professionals need to leverage existing knowledge from research on automation, such as, automation awareness and UI transparent design. For life-critical autonomous systems (e.g., autonomous vehicles and weapons), we advocate the implementation of the "tracking" and "tracing" mechanisms in system design of autonomous systems (de Sio & den Hoven, 2018), so that we can trace back the error data to identify the accountability of system versus human operators if the autonomous system fails and also be able to use the data for future improvement of the design (de Sio & den Hoven, 2018; Xu, 2021).

*Integration of HCI approaches into AI development*. Although multiple sets of ethical guidelines are currently available, the AI community lacks common aims, professional norms, methods to translate principles into practice. In many cases, ethical design was considered after the technology is developed rather than during development (Mittelstadt, 2019). Also, current ethical guidelines lack technical detail or detailed examples of AI ethical design options. How then can we embed ethical AI design principles within AI itself, increase the effectiveness of ethical codes with respect to behavior of AI engineers, and enhance AI engineers' skillset in ethical design with appropriate examples of ethical AI design? HCI professionals may offer support. For example, we can leverage human-centered HCI methods (e.g., iterative design and testing) to translate user needs into data needs, develop effective approach to generate, train, optimize, and test AI-generated behaviors during development, continuously improve the design to avoid biased outcomes. Designing explainable AI systems will also help develop ethical AI.

*An HCI multidisciplinary approach*. As a multidiscipline, we need to help AI professionals in ethical AI design. AI engineers typically lack formal training in applying ethics to design in their engineering courses and tend to view ethical decision making as another form of solving technical problems. The AI community now recognizes that the ethical AI design requires wisdom and cooperation from a multidisciplinary field extending beyond computer science (Li & Etchemendy, 2018). Many of the ethical AI issues need solutions from social and behavioral science, such as human privacy, human acceptance of AI systems, human needs for decision-making when using AI systems. The HCI community can leverage their interdisciplinary skills to assess the ethical related issues and help propose solutions by adopting methods of social and behavioral science from a broader sociotechnical systems perspective.

## 3.8 Summary of the opportunities for HCI professionals to enable HCAI

To sum up what have been discussed and proposed for future HCI work, Table 3

summarizes the new opportunities for HCI professionals, and the expected HCAI-driven design goals across the seven main issues. Specifically, the Opportunities for HCI Professionals column lists the future HCI work discussed earlier for addressing the new challenges identified, and the Primary HCIA Design Goals column specifically lists the primary HCAI design goals to be achieved across these opportunities.

**Table 3 Summary of the opportunities for HCI professionals**

| Main Issues (Sections 3.1- 3.7) | Opportunities for HCI Professionals (Sections 3.1- 3.7) | Primary HCAI Design Goals (Figure 1) |
|---|---|---|
| From expected machine behavior to potentially unexpected behavior (Section 3.1) | • Application of HCI approach for managing machine behavior<br>• Continued improvement of AI systems during behavioral evolution<br>• Enhancement of software testing approach<br>• Leveraging HCI approach in design | • Human-controlled AI |
| From interaction to potential human-AI collaboration (Section 3.2) | • Clarification of human and AI's roles<br>• Modeling human-AI interaction and collaboration<br>• Advancement of current HCI approach<br>• Innovative HCI design to facilitate human-AI collaboration | • Human-driven decision-making<br>• Human-controlled AI |
| From siloed machine intelligence to human-controlled hybrid intelligence (Section 3.3) | • Human-controlled hybrid intelligence<br>• Cognitive computing and modeling<br>• HCI framework for hybrid intelligence<br>• Human role in future human-computer symbiosis | • Augmenting human<br>• Human-controlled AI |
| From user interface usability to AI explainability (Section 3.4) | • Effective interaction design<br>• Human-centered explainable AI<br>• Acceleration of the transfer of psychological knowledge<br>• Sociotechnical systems approach | • Explainable AI |
| From human-centered automation to human-controlled autonomy (Section 3.5) | • Understanding the impacts of autonomous characteristics<br>• Innovative design paradigms<br>• Design for human-controlled autonomy<br>• A multidisciplinary HCI approach | • Human-controlled autonomy |
| From conventional interaction to intelligent interactions (Section 3.6) | • New interaction paradigms<br>• Usable user interface<br>• Adapting AI technology to human capability<br>• HCI design standards for AI systems | • Usable AI |
| From general user needs to specific ethical AI needs (Section 3.7) | • Adoption of meaningful human control in design<br>• Integration of HCI approaches into AI development.<br>• An HCI multidisciplinary HCI approach | • Ethical & responsible AI |

Our review and analyses in this section provide the answers to the second research question: What are the opportunities for HCI professionals to lead in applying HCAI to address the new challenges? As a result, our holistic assessment has enabled us to specifically identify the unique challenges as we transition to interaction with AI systems from the HCAI perspective; we also tie these new HCI opportunities to HCAI-driven design goals, eventually guiding HCI professionals addressing the new issues in developing AI systems.

We urge HCI professionals to proactively participant in the research and application of AI systems to exploit these opportunities for addressing the new challenges. Addressing the new challenges through these opportunities will promote the HCAI approach in developing AI systems, further advancing the HCAI approach and influencing the development of AI systems.

## 4. The need to improve HCI approaches for applying HCAI

This section is to provide the answers to the third research question: How can current HCI approaches be improved in the application of HCAI? Our analyses of current HCI approaches herein focus on the assessment of the current HCI methods being used and how HCI professionals currently influence the development of AI systems in terms of HCI processes and professional skillset; if gaps exist, how we should improve the current HCI approaches to enable the HCAI approach.

Existing HCI approaches were primarily defined for non-AI computing systems and may not effectively address the unique issues in AI systems. Future HCI work inevitably puts forward new requirements on these HCI approaches if we want to effectively address these new challenges by applying HCAI. The previous HCAI work has not included a comprehensive assessment of current HCI approaches (e.g., methods, process, skillset) on whether we have gaps to enable HCAI in future HCI work. Recent research already reported that many challenges have been encountered in the process of developing AI systems for HCI professionals (e.g., Yang et al., 2020).

## 4.1 Enhancing current HCI methods

Research shows that there are a lack of effective methods for designing AI systems and HCI professionals have had difficulty performing the typical HCI activities of conceptualization, rapid prototyping, and testing (Yang et al., 2020; Holmquist, 2017; Dove et al., 2017; van Allen., 2018). Driven by the unique characteristics of AI technology as discussed earlier, there are new needs that pose challenges to the existing HCI methods. For instance, the pervasive computing environment and ecosystems of artifacts, services, and data has challenged the existing HCI methods that focus on single user-artifact interaction (Brown et al., 2017).

The HCI community has realized the need to enhance existing methods (Stephanidis, Salvendy et al., 2019; Palvendo, 2019; Xu, 2018; Xu & Ge, 2018). To effectively address the identified unique issues of AI systems as discussed earlier, we need to enhance existing methods and leverage the methods from other disciplines. To this end, we assessed the over 20 existing methods of HCI, human factors, and other related disciplines from the HCAI perspective. As a result, Table 4 summarizes a comparison between the current HCI methods (i.e., typical HCI methods being used in designing non-AI systems) and the selected 7 alternative methods that are presented by enhancing existing HCI methods and leveraging the methods from other disciplines (e.g., Jacko, 2012; Endsley et al., 2012; Lieberman, 2009).

**Table 4   Comparison between conventional HCI methods and the alternative HCI methods (selected)**

| R & D stages of AI systems | Needs for HCI Professionals to Make Contributions in Developing AI Systems | Limitations of Conventional HCI Methods (e.g., Jacko, 2012) | Alternative HCI Methods (Selected) | Characteristics of the Alternative HCI Methods |
|---|---|---|---|---|
| User research, HCI test | Comprehensively assess the impacts of the entire pervasive computing environment and optimize the design of AI systems such as an intelligent Internet of Things (e.g., Oliveira et al., 2021) | Focus on a single user-computing artifact interaction with limited context of use | Scaled up & ecological method | Study the impacts of the entire pervasive computing environment (multiple users and AI agents) and the ecosystems of artifacts, services, and data in distributed contexts of use (Brown et al., 2017) |
| User research, HCI test | Comprehensively assess the impacts of AI technologies, and optimize the design to support people's daily work and life (e.g., Jun et al., 2021) | Limited to lab-based study, cannot effectively assess the broad impacts of AI on people's daily life and work | "In-the-wild" study | Carry out in-situ development and engagement, sampling experiences, and probing people in the field (e.g., home, workplace) to fully understand people's real experience and behavior while interacting with AI (Roger et al., 2017) |
| System and user needs analysis | Utilize the learning ability of AI systems to dynamically and intelligently replace more manual tasks and improve the overall performance of human-machine systems (e.g., Xu, Dov, et al., 2019) | Static and unchanging allocation of human-machine functions and tasks | Dynamic allocation of human-machine functions | Dynamic allocation of human-machine functions and tasks as intelligent machines learn over time, emphasizing the complementarity of human and machine intelligence |
| System analysis and design, human-machine functional analysis | Optimize the human-machine collaboration and performance of AI systems by taking advantage of the functional complementarity and adaptability between humans and AI systems | Machine works as a tool, basically no collaboration between human and machine | Human-machine teaming based collaborative design | Machine works as a tool + teammate; emphasize on the human-machine teaming relationship, shared information, goals, tasks, and autonomy between humans and AI systems (e.g., Johnson & Vera, 2019) |
| Low-fidelity prototyping, HCI test | At the early stage of development, prototype and test intelligent capabilities of AI systems to assess and validate design ideas (e.g., Martelaro, et al., 2017) | Focus on the non-intelligent functions, difficult to present intelligent functions | Prototyping of machine intelligent functions | Use Wizard of Oz (WOz) prototyping method to emulate and test intelligent functions of an AI system and design ideas at early stage (Martelaro, et al., 2017) |

| System design, prototype | AI/intelligence is used as a tool to truly empower HCI designers; technology becomes a valuable tool, facilitating designers throughout HCI iterative design and evaluation (Yang, 2018) | No tool to effectively help design AI systems, HCI professionals need to learn the technical details of AI | AI as a design material | Plug in AI/intelligence as a new design material in developing AI systems without having technical know-how (Holmquist, 2017) |
| --- | --- | --- | --- | --- |
| Needs analysis, system design | Provide personalized capabilities and contents based on real-time digital personas, user behaviors, and usage context (e.g., Kleppe, 2017) | Difficult to predict user needs, hard to obtain real-time data, such as user behaviors and contextual information | Big data-based interaction design | Model real-time user behaviors and contextual scenes using AI algorithms and big data to produce digital personas and user's usage scenarios, understand personalized user needs in real time (e.g., Berndt et al., 2017) |
| HCI evaluation | Assess AI systems and behaviors as AI systems evolve over time, optimize interaction design and potential human-AI collaboration from longitudinal perspective (Wang & Siau, 2018) | Limited to make interaction design decisions at a fixed time without considering the evolvement of AI-based machine behavior over time | Longitudinal study | Assess the performance and impacts of human-AI systems or interface as AI systems evolve over time, including potential human-AI collaboration (Lieberman, 2009) |

In Table 4, the Needs for HCI Professionals to Make Contributions in Developing AI Systems column describes what are required to develop HCAI systems, which is the gap in current HCI methods that are primarily applied to the development of non-AI systems (e.g., Jacko, 2012; Xu, 2017), as highlighted by the Limitations of Conventional HCI Methods column. The Characteristics of the Alternative HCI Methods column describes the characteristics of the selected 8 alternative methods based on the advantages of these methods advocated by the referenced researchers.

For instance, when at the developmental stage of determining requirements for an AI system, HCI professionals may use the following methods (among others):

- The 'scaled up & ecological method" and/or the "in-the-wild study" to assess the impacts of AI technology on users and gather user requirements from a broad perspective of sociotechnical systems

- The 'dynamic allocation of human-machine functions" method, the "big-data-based interaction design" method, and/or the "human-machine teaming based collaborative design" method to support HCI and system design

- The "AI as a design material" approach and the "WOZ design" approach to quickly prototype and test design concepts for the AI system to ensure the design meets user needs.

Next, they can follow up an iterative process to improve and test the design. Once the system is released to the market, HCI professionals may apply the "scaled up & ecological" method and/or the "in-the-wild study" method to further assess the AI system and its potential impacts on users. If necessary, they may adopt the "longitudinal study" method to assess the performance and impacts of the AI system as the system behavior evolves over time, including potential human-AI collaboration.

As shown in Table 4, these 7 alternative HCI methods can help HCI professionals overcome the limitations of conventional HCI methods when applying HCAI in developing AI systems. In addition, the advantages of these alternative HCI methods can be taken by HCI professionals for their HCI work across various development stages (see the R & D stages of AI systems column). Thus, we can draw our initial conclusion that there are gaps in current HCI methods in implementing HCAI, our comprehensive assessment helps identify alternative HCI methods for HCI professionals to be more effectively in developing AI systems to achieve the HCAI design goals in future HCI work. In addition, we encourage HCI professionals to leverage these alternative HCI methods and also expect these alternative methods will be improved in future HCI work.

**4.2 Influencing the development of AI systems**

As HCAI is an emerging approach, HCI professionals are also challenged by how we can effectively influence the development of AI systems when HCI methods are available. We argue that the effectiveness of the influence relies on several factors, including the integration of HCI methods into the development, HCI professionals' skillset and knowledge, acceptance of HCAI by AI professionals, and so on.

Research shows HCI professionals have challenges in integrating HCI processes into the development of AI systems. For instance, many HCI professionals still join AI projects only *after* requirements are defined, a typical phenomenon when HCI was an emerging new field (Yang, Steinfeld et al., 2020). Consequently, the HCAI approach and design recommendations from HCI professionals could be easily declined (Yang, 2018). AI professionals often claim that many problems that HCI could not solve in the past have been solved through AI technology (e.g., voice and gesture input), and they are able to design the interaction by themselves. Although some of the designs are innovative as developed by AI professionals, studies have shown that the outcomes may not be acceptable from a usability perspective (e.g., Budiu & Laubheimer, 2018). Also, some HCI professionals find it challenging to collaborate effectively with AI professionals due to lack of a shared workflow and a common language (Girardin & Lathia, 2017). Recent studies have shown that HCI professionals do not seem to be prepared themselves for providing effective design support for AI systems (Yang, 2018). We offer several strategic recommendations below on how

HCI professionals can effectively applying HCAI and influence the development of AI systems, which will further advance HCAI in future HCI work.

First, HCI professionals need to integrate the existing and enhanced HCI methods into the development process of AI systems to maximize the interdisciplinary collaboration. A multi-disciplinary team, including AI, data, computer science, and HCI, may bring various frictions and opportunities for misunderstandings, which must be overcome through process optimization (Girardin & Lathia, 2017; Holmquist, 2017). For instance, in order to understand the similarities and differences in practices between HCI professionals and the professionals from other disciplines, Girardin & Lathia (2017) summarize a series of touch points and principles. Within the HCI community, researchers have indicated how the HCI process should be integrated into the development of AI systems and the role that HCI professionals should play in the development of AI (Lau et al., 2018). For example, Cerejo (2021) proposed a "pair design" process that puts two people (one HCI professional and one AI professional) working together as a pair across the development stages of AI systems.

Second, HCI professionals should take a leading role to promote HCAI as we did 40 years ago for promoting the "human-centered design" approach for PC applications. In response to the challenges faced by the interdisciplinary communities such as human factors and HCI, past president of the Human Factors and Ergonomics Society (HFES), William Howell, put forward a model of "shared philosophy" by sharing the human-centered design philosophy with other disciplines (Howell, 2001), instead of a "unique discipline" model by claiming the sole ownership of the human-centered design by a discipline. Over the past 40 years, the participation of the professionals from HCI, human factors, computer science, psychology, and other disciplines into the fields of HCI is the embodiment of the "shared philosophy" model. In the early days of the AI era, it is even more necessary for HCI professionals to actively share the HCAI design philosophy with the AI community. Although the HCAI approach may not be fully accepted by all in the initial stage, but we need to minimize this time-lagging effect. Influence ultimately determines our continuing efforts, just as we jointly have been promoting user experience to society over the last 40 years and user experience has finally become a consensus for society.

Third, HCI professionals must update their skillset and knowledge in AI. Some HCI professionals have already raised their concerns (Yang, 2018). While AI professionals should understand HCI approaches, HCI professionals also need to have a basic understanding of AI technology and apply the knowledge to facilitate the process integration and collaboration. Opportunities include employees' continuing education, online courses, industry events and webinars, workshop, self-learning, or even learning through handling low-risk projects to develop new skills, so that HCI professionals can fully understand the design implications posed by the unique characteristics of AI systems and be able to overcome weaknesses in ability to influence AI systems as reported (Yang, 2018).

In a long run, we need to train our next generation of developers and designers to create HCAI systems. Over the past 40 years, HCI, human factors, and psychology disciplines have provided an extensive array of professional capabilities. These capabilities have emerged into a relatively mature user experience culture (Xu, 2005). For this to occur for HCAI, new measures at the level of university education are required. For instance, relevant departments or programs (e.g., HCI, Human Factors, Computer Science, Informatics, Psychology) need to proactively cultivate interdisciplinary talents doing HCAI work for society. More specifically, it is necessary to provide students with multiple options to learn interdisciplinary knowledge, such as, hybrid curriculums of "HCI + AI or HCAI," "AI major + social science minor," or "social science major + AI minor," and establish master and doctorate programs targeting HCAI.

Fourth, HCI professionals need to proactively initiate applied research and application work through cross-industry and cross-disciplinary collaborations to increase their influences. Scholars in academia should actively participate in AI related collaborative projects across disciplines and collaboration between industry and academia. The development of autonomous vehicles is a good example. Many companies currently have heavily invested in developing autonomous vehicles. As discussed earlier, there are opportunities for academia to partner with industry to overcome the challenges. Also, the human-machine hybrid augmented intelligence, as advocated by the HCAI approach, needs collaborative work between academia and industry.

Finally, we need to foster a mature environment for implementation of HCAI, including government policy, management commitment, organizational culture, development process, design standards and governance, and development process. We firmly believe that a mature HCAI culture will eventually come in, the history has proved our initial success in promoting HCI and user experience in the computer era. The HCI community needs to take the leading role again.

## 5. Conclusions

This paper has answered three research questions: (1) What are the challenges to HCI professionals to develop human-centered AI systems (Section 2); (2) What are the opportunities for HCI professionals to lead in applying HCAI to address the challenges (Section 3); and (3) How can current HCI approaches be improved to apply HCAI (Section 4)? The primary contributions of this paper and the urgent messages to the HCI community as actions are as follows.

Driven by AI technology, the focus of HCI work is transitioning from human interaction with non-AI computing systems to interaction with AI systems. These systems exhibit unique characteristics and challenges for us to develop human-centered AI systems. While they have benefited humans, they may also harm humans if they are not appropriately developed. We have further advanced the HCAI approach specifically for HCI professionals to take actions. The HCI community must fully understand these unique challenges and take the human-centered AI approach to maximize the benefits of AI technology and avoid risk to humans.

Based on the HCAI approach and the specified HCAI-driven design goals, we have identified new opportunities across the seven main issues that the HCI community needs to take a leading role to effectively address human interaction with AI systems. The emergence of these unique AI related issues is inevitable as technology advances, just as it was during the emergence of HCI in the 1980's. Today, we are just facing a new type of machine - AI systems that present new challenges to us, compelling us to take action by taking the HCAI approach.

This paper has identified main gaps in current HCI approaches in applying HCAI in order to effectively address the new challenges and opportunities. We call for action by the HCI community. Specifically, we need to:

- Enhance and integrate HCI methods into the development of AI systems to maximize the interdisciplinary collaboration; take the leading role to promote the HCAI approach

- Update our skillsets and knowledge in AI and train the next generation of developers/designers for developing HCAI systems

- Proactively conduct applied research related to AI systems through cross-disciplinary collaboration

- Foster a mature organizational environment for implementing the HCAI approach.

History has proven through our initial success in promoting human-centered design in the PC era in the past, the HCI community needs to take the leading role again.

## Acknowledgements

The authors appreciate the insightful comments from Professor Ben Shneiderman and four anonymous reviewers on an earlier draft of this paper. These insights have significantly improved the quality of this paper. Any opinion herein is those of the authors and does not reflect the views of any other individuals or corporation.